# Comment on "Spectroscopic Evidence for Multiple Order Parameters in the Heavy Fermion Superconductor CeCoIn$_5$"

**Goutam Sheet and Pratap Raychaudhuri**
*Tata Institute of Fundamental Research, Homi Bhabha Road, Mumbai-400005*
*India*

In ref.1, the authors report PCS studies of the heavy Fermion superconductor CeCoIn$_5$. They have reported two kinds of spectra (fig.1a and fig. 1b) obtained using Pt-Ir tip. Fig.1a of ref.1 shows a zero bias conductance peak followed by two dips appearing symmetrically about the central peak. Fig. 1b of the same paper shows an asymmetric hump structure. They interpret the origin of these spectra as the presence of Andreev surface states and Andreev bulk states. They have taken the central peak as the signature of existence of nodes in the order parameter. They attempted to analyze the spectra using d-wave BTK formalism and obtained support for multiband superconductivity in super conducting CeCoIn$_5$.

However, in point contact experiments, those spectral features are regularly observed in a variety of contacts between different combinations of a normal metal and a superconductor. In reference 2, the origin of the dip structures have been extensively discussed when the contact is not in the ballistic limit. In ref. 1 the authors have calculated the contact diameter as 70 nm, which is 87.5% of the mean free path (80nm) reported inside this sample. Again, while calculating the contact diameter they have equated the normal state resistance with the sharvin resistance only, ignoring the contribution from Maxwell's resistance. If the Maxwell's contribution is also added, the contact diameter will be even more, and possibly the reported spectra were obtained with contacts close to the thermal regime. Therefore, BTK formalism is not applicable here to analyze the spectra.

An example of how the same combination of sample and tip can give rise to both "canonical" spectrum as well as unconventional spectrum depending on the contact size is shown in Figure 1. Figure 1(a) shows a canonical spectrum for a Fe foil/.Nb-tip contact. Figure 1(b) shows the another spectrum using the same combination of sample and tip showing all the features which Rourke-et al. interpret as evidence for unconventional superconductivity. The evolution of the spectra with junction impedance is discussed in ref. 2 and also evident from fig. 2. In fig. 2, the dips in the point contact spectra between a single crystal of YNi$_2$B$_2$C and Au (tip) vanishes systematically with increasing contact resistance. The dip structures may or may not appear for non-ballistic point contacts depending on the critical current of the contact geometry. Again, the hump structures mentioned in ref. 1 are also often obtained for several superconductors. Fig. 1(b) shows one spectra showing both the hump and the dip structures (marked by arrows) for a point contact between Fe and Nb. In fact, we have observed the dips and hump structures for several point contacts between a normal metal and a conventional superconductor, a ferromagnet and a conventional super conductor, a normal metal and an unconventional superconductor.

The spectra obtained from point contact experiments between a normal metal and a super conductor should be chosen carefully before giving attempt to analyze them using BTK theory. Spectra with many interesting but so far inexplicable features can be obtained depending on the contact geometry. Since it is difficult to control the contact geometry, they are not always reproducible. To analyze them, one should have the detail knowledge about the contact geometry. Sometimes, the contact can be made at multiple points. It is nearly impossible to consider all the parameters involved to analyze ***all types of spectra*** to obtain reliable quantitative values of the order parameter and conclude anything about its symmetry in practical life.

However, fortunately, there are some known spectral features, which can be comfortably analyzed using BTK theory. A spectrum with two peaks at gap voltage symmetric about V=0 (shown by arrows in fig. 1(a)) can be taken as a good spectra if dips do not appear at higher voltages. Again, if the zero bias conductance peak undergoes Zeeman splitting in the presence of magnetic field then it can be considered as the signature of the presence of Andreev bound states. Apart from spectra having these two features, it is not trivial to conclude anything from point contact experiments between a normal metal and a super conductor.

In summary, according to our observation, the PCS spectra shown in ref. 1 are obtained for contacts close to the thermal regime. Therefore, they cannot be taken as support for the unconventional superconductivity and multiple order parameters in $CeCoIn_5$.

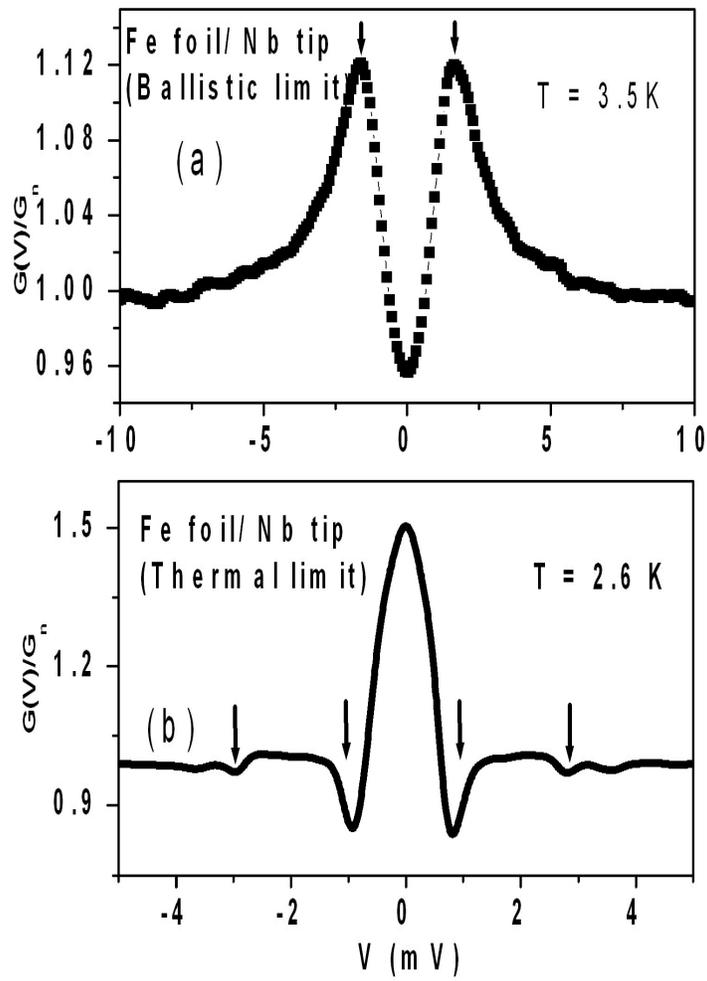

Figure 1

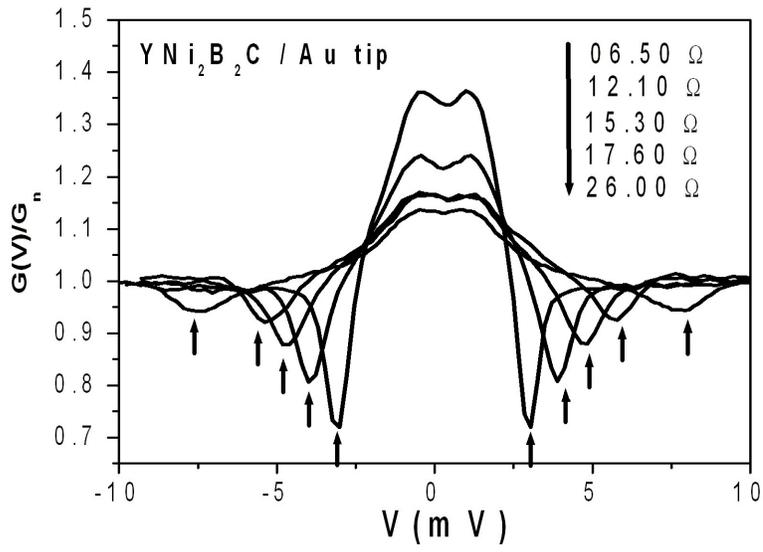

Figure 2